\documentclass[12pt]{iopart}
\begin{document}
\title{Radiation reaction of a classical quasi-rigid extended
 particle.}
\author{Rodrigo Medina}
\address{Instituto Venezolano de Investigaciones Cient\'{\i}ficas, 
Apartado 21827, Caracas 1020A, Venezuela}
\ead{rmedina@ivic.ve}

%%%%% Abstract %%%%%%
\begin{abstract}
The problem of the self-interaction of a quasi-rigid classical
particle with an arbitrary spherically symmetric charge distribution is
completely solved up to the first order in the acceleration. No 
\textit{ad hoc} assumptions are made. The relativistic
equations of conservation of
energy and momentum in a continuous medium are used. The electromagnetic fields
are calculated in the reference frame of instantaneous rest using the Coulomb
gauge; in this way the troublesome power expansion is avoided.  Most
of the puzzles that this problem has aroused are due to the inertia of the
negative pressure that equilibrates the electrostatic repulsion inside
the particle. The effective mass of this pressure is $-U_\rme/(3c^2)$, where
$U_\rme$ is the electrostatic energy. When the pressure mass is taken into
account the dressed mass $m$ turns out to be the bare mass plus the
electrostatic mass $m=m_0+U_\rme/c^2$. It is shown that a proper mechanical
behaviour requires that $m_0>U_\rme/(3c^2)$. This condition poses a lower
bound on the radius that a particle of a  given bare mass and charge may
have. The violation of this condition is the reason why the
Lorentz-Abraham-Dirac formula for the radiation reaction of a point
charge predicts unphysical motions that run away or violate causality.
Provided the mass condition is met the solutions of the exact equation of motion
never run away and conform to causality and conservation of energy and momentum.
When the radius is much smaller than
the wave-length of the radiated fields, but the mass condition is still met,
the exact expression reduces to the formula that Rohrlich [\textit{
Phys. Lett. A} \textbf{303} 307 (2002)] has advocated for
 the radiation reaction of a quasi-point charge.
\end{abstract}
\submitto{\JPA}
\maketitle

%%%%% Section I
\section{Introduction}

Perhaps the last unsolved problem of classical physics is the self-interaction
of charged particles. We refer to a classical particle with a
spherically symmetric charge density moving under the influence of an
external force $\bi{F}_{\rm ex}$ and the force $\bi{F}_{\rm em}$
due to the electromagnetic  fields that it itself
generates. This last self-force can be expressed as the sum of the negative
time derivative of the momentum of the local electromagnetic fields that
surround the particle ($\bi{F}_{\rm L}$) plus the radiation reaction 
($\bi{F}_{\rm R}$). Fairly complete relations of  more than a century of
history can be found in the monograph of Yaghjian \cite{Yaghjian}, in the
review of Rohrlich \cite{Rohrlich1} and in the chapter 16 of the textbook of
Jackson \cite{Jackson}. The published solutions of this problem
are filled with puzzles. Let us consider the most notorious ones.
We will use Gauss units.

 1){\it Spontaneous self-acceleration}.
The Lorentz-Abraham-Dirac (LAD) formula for the radiation reaction of a point
charge \cite{Lorentz,Abraham1,Dirac}, 
which in the non-relativistic limit reduces to 
$\bi{F}_{\rm R}=\frac{2}{3} q^2\dot\bi{a}/c^3$, has solutions in which
the particle, in 
the absence of an external force, spontaneously accelerates itself 
approaching speed $c$ (run-away solutions).

 2){\it Violation of causality}. There
are also solutions of the LAD equation that do not run away, but where
the acceleration depends on values that the applied force will take in the
future.

 3){\it Power unbalance}. The electrostatic field ({\it i.~e.}
the electric field that is present when the particle is not accelerated) does
not contribute to the net force on the particle, because the integral over the
volume of the corresponding force density vanishes. It, nevertheless,
contributes to the power, because of the Lorentz contraction. As a result
the total power
of electromagnetic forces differs from the product
 $\bi{F}_{\rm em}\!\cdot\bi{v}$. This was noticed by Abraham in
 1904 \cite{Abraham2}. In 1905 Poincar\'e \cite{Poincare1,Poincare2}
showed that the existence of a pressure was required in order to balance the
electrostatic repulsion and thus assure the stability of the particle. The work
of the pressure as the volume of the particle changes,
was supposed to fix the power unbalance. But
there is a problem, in Ref. \cite{Poincare1} and in the introduction of
~Ref.  \cite{Poincare2} (p. 20) Poincar\'e  refers to an {\it external}
pressure, whose work is indeed proportional to the change of volume, but
the pressure that is actually calculated (p. 63 of Ref.\ \cite{Poincare2}) is
 an {\it internal} negative
pressure, whose work over the whole particle is obviously zero.
It is worth remarking that the results of Lorentz, Abraham and Poincar\'e on
this subject conform with relativity in spite of being previous to the 1905
Einstein paper, because they assume an \textit{ad hoc} Lorentz contraction.

4) {\it The $4/3$ problem}. The
electrostatic energy $U_\rme$ should contribute to the mass of the particle, so
if $m_0$ is the bare mass then the dressed mass must be $ m= m_0+U_\rme/c^2$.
 But if one computes the momentum of the electromagnetic field that surrounds
a spherical charge distribution moving with velocity $\bi{v}$ one obtains
$ \frac{4}{3}U_\rme\gamma \bi{v}/c^2$. This value agrees with that part of the
self-force that is proportional to the acceleration ($\bi{F}_{\rm L}$).
 To make matters more
confusing the energy of such  field is $U_\rme\gamma(1+\frac{1}{3}(v/c)^2)$,
which has an additional $\frac{1}{3}(v/c)^2$ term.

Schwinger \cite{Schwinger} has shown,  for a spherically symmetrical
charge distribution moving uniformly, that there are tensors which depend on
the fields, that can be added to the electromagnetic energy-momentum-stress
tensor $T^{\mu\nu}$ in order to obtain a divergence-less total tensor.  He
proposed that such ``corrected'' tensors should be used for calculating the
four-momentum of the fields. 
 He found two
possible tensors, corresponding to  electromagnetic masses $U_\rme/c^2$ and
 $\frac{4}{3}U_\rme/c^2$. Rohrlich \cite{Rohrlich1} has remarked that any value
of the electromagnetic mass in between is also possible.

Instead Singal \cite{Singal} has shown that the conventional theory
is consistent when a proper account is taken of all the energy and momentum
associated with the electromagnetic phenomenon in the system.

5){\it Apparent violation of energy and momentum conservation}. The Larmor
\cite{Larmor} formula for the radiated power 
by a point charge in the instantaneous rest
frame  is $P_{\rm rad}=\frac{2}{3}q^2a^2/c^3$. But the LAD formula
 predicts that if the particle is moving with
 constant acceleration the reaction force is
zero and therefore also its power. From where is the radiated power coming?
 A similar situation occurs with the momentum. In the rest frame the rate of
radiated momentum vanishes. Where does the impulse of the radiation reaction
force go? One should expect that the electromagnetic field surrounding the
particle acts as a reservoir of energy and momentum.

 The Classical Electromagnetism of a continuous current density $j^{\mu}$ is a
perfectly well behaved theory  that conserves energy and momentum and
conforms to causality. Solutions that run away or violate
causality simply cannot appear. Various are the sources of the problems.

 The run-away solutions and the violations of causality should be attributed
 to the $R \rightarrow 0$ limit. We will show for any spherical charge
distribution, that in order to avoid run-away solutions the total effective
mass of the particle $m$ must be greater than the mass of the
electromagnetic fields surrounding it, {\it i.~e.} $\frac{4}{3}U_\rme/c^2$.
 The reason is
 simple: $m-\frac{4}{3}U_\rme/c^2$ is the effective mass of {\it matter}, if
it  were negative the acceleration would be opposite to the force.
 As $U_\rme\geq\frac{1}{2}q^2/R$ there is an absolute minimum for the radius
of a classical particle of mass $m$ and charge $q$,
 $R\geq\frac{2}{3}q^2/(mc^2)$. Therefore the $R \rightarrow 0$ limit is
unphysical, because $R=0$ is interior to an unphysical region.

The violation of causality is due to the expansion in powers of $R$ that
is used to evaluate the electromagnetic fields and the self-force. As the
fields are linear functionals of the {\it retarded} currents $j^\mu$, so must
also be $\bi{F}_{\rm em}$. As this self-force vanishes if the particle is
not accelerated, it should be a functional of the retarded acceleration
$\bi{a}(t-|\bi{x}-\bi{y}|/c)$, where $\bi{x}$ and $\bi{y}$ are
two points of the particle. We will find such functional for an arbitrary
charge density. So, for the exact solution of an extended particle there
are no pre-accelerations or violations of causality. Of course if we make
an expansion in powers of $|\bi{x}-\bi{y}|$ and we take any finite
number of terms we get an instantaneous result which violates causality.
If we take the $R \rightarrow 0$ limit we are left with the $R^{-1}$ and
the $R^0$ terms, the first one diverges and corresponds to
the local fields reaction $\bi{F}_{\rm L}$, while
the second is the LAD result for the radiation reaction $\bi{F}_{\rm R}$;
as said above such limit is unphysical. In the calculation of
Dirac \cite{Dirac} the
violation of causality is explicit, because the {\it advance} Green's
function is used.

 The exact self-force of a sphere of radius $R$ with a
uniformly charged surface has been calculated by
 Sommerfeld \cite{Sommerfeld}, Page \cite{Page}, Caldirola \cite{Caldirola} and
Yaghjian \cite{Yaghjian}. Their result confirm what we have been
 discussing \cite{Rohrlich1,Rohrlich2}.
 Our general result yields the Sommerfeld-Page-Caldirola-Yaghjian  formula
 for the charged spherical shell.

The power unbalance and the 4/3 problems have a common origin. They are due
to having overlooked a purely relativistic effect:
{\it pressure has inertia}.
In our calculation we use the relativistic momentum-energy conservation
formula for a continuous medium. Here the momentum-energy tensor of matter,
the stress tensor and the force density appear. We have to remark that
this stress tensor is not some {\it ad hoc} addition, it is just the stress
tensor that any continuous medium should have. At rest the stress tensor
 is determined by the electrostatic
repulsion and is purely spatial, but when the
particle is moving the temporal components of the tensor do not vanish.
Those components  contribute to the momentum density and to the energy
density of the particle. By integration over the volume one obtains the
equation of motion and the energy equation. The pressure contribution
to the mass of the particle is $-\frac{1}{3}U_\rme/c^2$ which is just what
is needed to fix the 4/3 problem. The power unbalance also disappears.

Finally the apparent violation of energy and momentum conservations is
understood by considering all the contributions to the energy and
momentum of the electromagnetic field that surrounds
the particle. When the particle is not accelerated there is an electromagnetic
field around the particle, the local field, which contributes
to the energy and momentum of the particle. Call $u_{\rm L}$
and $\bi{S}_{\rm L}$
its energy density and Poynting vector. In addition to this, when the
particle is accelerated there is also the radiated field with energy
density $u_{\rm R}$ and Poynting vector $\bi{S}_{\rm R}$. But the energy density
and the Poynting vector are quadratic functions of the fields, so the 
total energy density and Poynting vector have additional cross terms,
$\bi{S}=\bi{S}_{\rm L}+\bi{S}_{\rm R}+\bi{S}_{\rm C}$ and
 $u=u_{\rm L}+u_{\rm R}+u_{\rm C}$ These
cross terms act as energy and momentum reservoirs.

The point charges ($R=0$) are inconsistent with classical electrodynamics, but
phenomenologically a point charge is a body with a radius that is
negligible as compared to the other relevant distances, in particular
to the
wavelength of the  fields with which it interacts. We will call such
particles  physical point charges, while we call mathematical 
point charges those with $R=0$. We will show that
the acceleration can be expressed as a time integral of the retarded
applied force. When the physical point charge conditions are fulfilled the
expression reduces to an equation similar to LAD motion equation, but
replacing $\dot\bi{a}$ by $\dot\bi{F}_{\rm ex}/m$, exactly as was
recently proposed by Rohrlich \cite{Rohrlich3}.

%%%% Section II

\section{Quasi-rigid motion}
A perfectly rigid  body is inconsistent with relativity because it
implies instantaneous propagation of signals. In general if a body is
accelerated it is stressed and deformed. How this happens depends on
the particular characteristics of the body and the applied force.
If the external force is more o less uniform through the body and
the acceleration is small and does not change too rapidly the body
can, as seen in the instantaneous rest frame, continuously maintain its
shape suffering a negligible deformation compared with particle radius.
The bound on the acceleration $\bi{a}$ is easily determined. 
The speed change during the time that a signal takes to
traverse the particle should be negligible with respect to the speed of
sound (speed of signals in the particle). So, $(2R/v_s)a\ll v_s$. For
a particle of maximum stiffness;  $v_s=c$ and so $a\ll c^2/(2R)$, which
is an immense bound even for macroscopic particles. The bound of the time
derivative of the acceleration is obtained by comparison of
the relative rate of change of the acceleration with the inverse of the
transit time, $\dot{a}/a\ll v_s/(2R)$. Again assuming  maximum 
stiffness, $\dot{a}/a\ll c/(2R)$.
 If both conditions are satisfied the motion of the particle is
essentially rigid and the internal degrees of freedom of the particle
are not excited. We will assume that this is the case. We are explicitly
excluding effects produced by deformation of the particle, such as polarisation.

Consider a spherically symmetric particle. Let $\bi{x}_{0}(t)$ be the position
of the centre of the particle in the laboratory reference frame. Let
$\bi{v}_{0}(t)$ and $\bi{a}_{0}(t)$ be the velocity and acceleration of
that point. The
instantaneous rest frame  would be defined by a Lorentz transformation of
velocity $\bi{v}_0$ and ``parallel'' axes. Let $\bi{y}$ be the position
relative to the centre of an element of the particle in the rest frame. We will
assume that the motion of each element of the particle is given by the
following expression
\begin{equation}\label{QuasiRigid}
\bi{x}(\bi{y},t)=\bi{x}_{0}(t) + \bi{y} + (\gamma_0^{-1}-1)(\bi{y}
\cdot\hat{\bi{v}}_0)\hat{\bi{v}}_0 ,
\end{equation}
were $\gamma_0=(1-(v_0/c)^2)^{-\frac{1}{2}}$. This expression is equivalent
to the Lorentz transformation of the laboratory relative position
$\bi{x}(\bi{y},t)-\bi{x}_{0}(t)$ to the rest frame position $\bi{y}$.
If the particle is not accelerated 
(\ref{QuasiRigid}) reproduces correctly the Lorentz contraction.

 If the particle is accelerated
the velocity is different in different points of the particle
\begin{eqnarray}
\bi{v}(\bi{y},t)&=\frac{\partial\bi{x}(\bi{y},t)}{\partial t}
\\
&=\bi{v}_{0}(t)+\frac{\rmd\gamma_{0}^{-1}}{\rmd t}
(\bi{y} \cdot \hat{\bi{v}}_0)\hat{\bi{v}}_0+(\gamma_{0}^{-1}-1)
\frac{\rmd}{\rmd t}[(\bi{y}\cdot\hat{{\bi{v}}}_0)\hat{\bi{v}}_0] ,
\end{eqnarray}
but in the rest reference frame all the points are at rest
 $\bi{v}(\bi{y},t)=0$.  The
accelerations in different points differ even in the rest frame, but the
difference in that case is negligible
 $\bi{a}(\bi{y},t)=(1-\bi{y}\cdot\bi{a}_0/c^2)\bi{a}_0$.

It is convenient to define $\delta\bi{v}$ as the difference between
the velocity of a point and the velocity of the centre
\begin{equation}\label{deltav}
\delta\bi{v}=\bi{v}(\bi{y},t)-\bi{v}_0(t) .
\end{equation}

%%%%%% Section III

\section{Extended particle equation}
The dynamics of an extended particle is determined by the momentum-energy
tensor $\Theta^{\mu \nu}$,  the stress tensor $P^{\mu \nu}$ and the force
density $f^{\mu}$ with the equation \cite{Bergmann}
\begin{equation}\label{ParExRel}
\nabla_{\mu}(\Theta^{\mu\nu}+P^{\mu\nu})=f^{\nu} .
\end{equation}

 For a spherically symmetric particle the proper density of
rest mass $\mu(y)$, which is a four-scalar, is a function of the
radial $y=|\bi{y}|$ coordinate in the rest frame. The integral of $\mu(y)$
over the proper volume is the bare rest mass of the particle $m_0$.
 The tensor $\Theta^{\mu \nu}$ is given by 
\begin{equation}
\Theta^{\mu \nu}=\mu u^{\mu}u^{\nu} ,
\end{equation}
where $u^{\mu}$ is the four-velocity ($u^i=\gamma v^i$,$u^0=\gamma c$).

The stress tensor in the rest frame is symmetric and purely spatial,
$P^{0\nu}=P^{\nu 0}=0$. We are using the metric tensor $g^{\mu \nu}$ that has
a positive trace. In general $P^{00}=v_iv_jP^{ij}/c^2$ and
$P^{0i}=P^{i0}=P^{ij}v_j/c$. At rest, because of  spherical
symmetry, the stress tensor must be of the form 
${\bf P}= P_r\hat{\bi{r}}\hat{\bi{r}}+
P_t({\bf I}-\hat{\bi{r}}\hat{\bi{r}})$, where ${\bf I}$ is the
identity tensor and $\hat{\bi{r}}$ is the radial unit vector. Also at rest
the force density $\bi{f}$ is the electrostatic repulsion and the
(\ref{ParExRel}) reduces to $\nabla\cdot{\bf P}=\bi{f}$. This
equation determines ${\bf P}$, but the relation between the radial
stress $P_r$ and the transverse stress $P_t$ depends on the particular
elasticity properties of the particle.  Nevertheless, the final result
after integrating over the volume, should be independent of the particular
elasticity model. Therefore,
 for the sake of simplicity we will assume
a Pascalian (isotropic) pressure; that is, in the rest frame
 $P^{ij}=P\delta_{ij}$. The pressure $P$ is a four-scalar. Then in a generic
reference frame
\begin{equation}\label{StressTensor}
P^{\mu \nu} = P(g^{\mu \nu} + u^{\mu}u^{\nu}/c^2) .
\end{equation}

The momentum and energy equations are then
\begin{equation}\label{ParExMomentum}
\frac{\partial}{\partial t}[(\mu+P/c^2)\gamma^2{\bf v}] + \nabla\cdot
[(\mu+P/c^2)\gamma^2\bi{v}\bi{v}]+\nabla P=\bi{f}
\end{equation}
and
\begin{equation}\label{ParExEnergy}
\frac{\partial}{\partial t}[(\mu c^2+P)\gamma^2 -P]+\nabla\cdot
[(\mu c^2+P)\gamma^2\bi{v}]=\bi{f}\cdot\bi{v} .
\end{equation}

It is interesting to compare these equations with the non-relativistic ones.
The only pressure terms that appear in the non-relativistic equations are
the gradient in (\ref{ParExMomentum}) and the pressure that is inside
the divergence of (\ref{ParExEnergy}). Note that the $P/c^2$ terms
in (\ref{ParExMomentum}) do not disappear in
the limit $ v/c \to 0$. This means that, in this case,
the usual non-relativistic limit, $ v/c \to 0$,
does not yield the actual non-relativistic result. In other
words, the inertia of pressure is
a purely relativistic phenomenon.

 Neglecting the inertia of pressure
is at the origin of some of the discussed puzzles. The $P/c^2$ terms that add
up to $\mu$ are related to the $4/3$ problem. The additional $-P$  term in the
time derivative of (\ref{ParExEnergy}) is related to the power
unbalance.

In order to obtain the equations of motion for the particle we integrate
over the whole volume. As $\mu$ and $P$  vanish outside the particle
the divergence terms disappear. That is the reason why, in the  Newtonian
mechanics, the pressure has no effect on the motion of the whole particle.
 Let us consider the first term of the 
left hand side of (\ref{ParExMomentum}),
\begin{equation}\label{Integ1}
\int\! \rmd^3x\,\frac{\partial}{\partial t}[(\mu+P/c^2)\gamma^2\bi{v}]=
\frac{\rmd}{\rmd t}\int\! \rmd^3x\,(\mu+P/c^2)\gamma^2(\bi{v}_0+\delta\bi{v})
\end{equation}
\begin{eqnarray}
\label{Integ2}
&=&\frac{\rmd}{\rmd t}\big[\gamma_0^2\int \rmd^3x\,(\mu+P/c^2)(\bi{v}_0
+\delta\bi{v}) \big]+ \Or(a^2)
\\
\label{Integ3}
&=&\frac{\rmd}{\rmd t}\big[\gamma_0^2\int \rmd^3x\,(\mu+P/c^2)\bi{v}_0\big]+
 \Or(a^2)
\\
\label{Integ4}
&=&\frac{\rmd}{\rmd t}\big[\gamma_0\int \rmd^3y\,(\mu+P/c^2)\bi{v}_0\big]+
\Or(a^2) .
\end{eqnarray}

In (\ref{Integ1}) we have used (\ref{deltav})
In (\ref{Integ2}) we have replaced $\gamma$ by $\gamma_0$,
$\gamma=\gamma_0+\Or(a)$. In (\ref{Integ2}) the  term of
$\delta\bi{v}$ vanishes because $\delta\bi{v}$ is odd in $\bi{y}$.
In (\ref{Integ4}) we have transformed the integral to the rest frame,
which gives a factor $\gamma_{0}^{-1}$.At the end we are
left with
\begin{equation}\label{ParMomentum}
\frac{\rmd}{\rmd t}[(m_0+m_{\rm P})\gamma_0\bi{v}_0] = \bi{F}
\end{equation}
where we have defined the pressure mass $m_{\rm P}$ and the 
total force $\bi{F}$ as
\begin{equation}
m_{\rm P}=c^{-2}\int \rmd^3y\, P
\end{equation}
and
\begin{equation}
\bi{F}=\int \rmd^3x\, \bi{f} .
\end{equation}

Similarly,
\begin{equation}\label{ParEnergy}
\frac{\rmd}{\rmd t}[(m_0+m_{\rm P})c^2\gamma_0 -m_{\rm P} c^2\gamma_0^{-1}] =
\bi{F}\cdot\bi{v}_0 + \int \rmd^3x\, \bi{f}\cdot\delta\bi{v} .
\end{equation}

Because of the following identity
\begin{equation}\label{identity}
c^2\frac{\rmd\gamma_0}{\rmd t}=
\bi{v}_0\cdot\frac{\rmd\gamma_0\bi{v}_0}{\rmd t}
\end{equation}
if $m_{\rm P}$ is constant, it must be that
\begin{equation}\label{PowerBalance}
-m_{\rm P} c^2\frac{\rmd\gamma_0^{-1}}{\rmd t}=
\int \rmd^3x\,\bi{f}\cdot\delta\bi{v} .
\end{equation}

The pressure has two effects: First it contributes to the effective
mass of the particle. Second, as long as (\ref{PowerBalance}) is
verified, it cures the power unbalance. The pressure mass $m_{\rm P}$ could
be negative, as it is in the charged particle case. A proper mechanical
behaviour requires a positive effective mass $m_0+m_{\rm P} >0$, otherwise the
acceleration would be opposite to the force.

 In the
rest of the paper there will be no confusion between the velocity of
a point in the particle and the velocity of its centre, so
in the following we will drop the zero subscript of $\bi{v}_0$,
$\bi{a}_0$ and $\gamma_0$.

%%%% Section IV

\section{Stresses in the charged particle}
Consider a particle with a spherically symmetric charge density, that
in the rest frame is $\rho (y)=qg(y)$, where $q$ is the charge.
The fraction $Q(r)$ of charge contained inside a sphere of radius $r$ is
defined as
\begin{equation}\label{Q}
Q(r)=4\pi \int^{r}_{0} \rmd y\,y^2g(y) ;
\end{equation}
of course
\begin{equation}
\frac{\rmd Q}{\rmd y}=4\pi y^2g(y)
\end{equation}
and $g(y)$ is normalised to $1$,
\begin{equation}
\int \rmd^3y\,g(y)=\lim_{y\to\infty}Q(y)=1 .
\end{equation}
The convergence of the integral requires
\begin{equation}
\lim_{y\to\infty}y^3g(y)=0 .
\end{equation}

 It is also required that the electrostatic energy be finite, this
implies
\begin{equation}
\lim_{y\to 0}y^{5/2}g(y)=0
\end{equation}
and
\begin{equation}
\lim_{y\to 0}y^{-1/2}Q(y)=0 .
\end{equation}

 The force density is the sum of an external force
density and the contribution of its own electromagnetic fields,
$\bi{f}=\bi{f}_{\rm ex} + \bi{f}_{\rm em}$, with
$\bi{f}_{\rm em}=\rho \bi{E}+ c^{-1}\bi{j}\times \bi{B}$.
The electromagnetic field $F^{\mu\nu}$ has two components, a local field
$F^{\mu\nu}_{\rm L}$ which decays as $r^{-2}$ and the radiation
field $F^{\mu\nu}_{\rm R}$
which decays as $r^{-1}$ and is proportional to the acceleration.
The self-force $\bi{f}_{\rm em}$ has therefore four terms corresponding
to the electric and magnetic forces of the local and radiation fields. The
local electric field depends on acceleration $\bi{E}_{\rm L}(\bi{a})$. We
will define the electrostatic force density as
 $\bi{f}_{\rm es}=\rho\bi{E}_{\rm L}(0)$ ; this is the largest term, by
definition it is independent of acceleration,
 but by symmetry it does not contribute to
the net force. It does contribute
to the power. The remaining local electric force density
$\rho(\bi{E}_{\rm L}(\bi{a})-\bi{E}_{\rm L}(0))$ and the
other three  terms will be included in the field reaction
force density $\bi{f}_{\rm fr}$,  so 
 $\bi{f}_{\rm em}= \bi{f}_{\rm es}+\bi{f}_{\rm fr}$.
The force of the local magnetic field, of course does not contribute to the
power, but it does give a contribution to the total force which is of first
order in the acceleration.

 The momentum
and energy of the local field are bound to the particle and contribute to
its total energy and momentum. When the particle moves with constant velocity
the local field is easily obtained with a Lorentz transformation of the
electric field at rest. When the particle is accelerated the local field
is modified far away from the particle. So the local field momentum and energy
are only slightly dependent on the acceleration. The zeroth order can be
calculated using the constant velocity fields.

 At rest the electric field is
\begin{equation}
\bi{E}=q \frac{Q(y)}{y^2}\hat{\bi{y}} .
\end{equation}

The electrostatic energy $U_\rme$ can be evaluated with any of the following
equivalent expressions
\begin{eqnarray}
U_\rme&=&\frac{q^2}{8\pi}\int \rmd^3y\,\frac{Q(y)^2}{y^4}
\\
&=& q^2\int_0^{\infty} \rmd y\,\frac{\rmd Q}{\rmd y}\frac{Q(y)}{y}
\\
&=&\frac{q^2}{2}\int\!\!\!\int
 \rmd^3y\, \rmd^3y\prime\,\frac{g(y)g(y\prime)}{|\bi{y}-\bi{y}\prime|} .
\end{eqnarray}

At rest the (\ref{ParExMomentum}) reduces to
 $\nabla P=\bi{f}_{\rm es}$, from where the pressure can be obtained.
In general it also depends on $\bi{f}_{\rm ex}$ and on
 $\bi{f}_{\rm fr}$, but as $f_{\rm fr}\ll f_{\rm es}$ and
as it is supposed that the external force does not deform the particle, those
dependencies will be neglected. In the frame of instantaneous rest the
pressure, up to zeroth order in the acceleration, is then
\begin{equation}\label{Pressure}
P(r)=-q^2\int_{r}^{\infty}\rmd y\, \frac{g(y)Q(y)}{y^2}+\Or(a) .
\end{equation}
By integration of this expression  over the volume, the pressure mass
is obtained,
\begin{equation}
m_{\rm P} =-\frac{U_\rme}{3c^2}+\Or(a) .
\end{equation}

When the particle moves with constant velocity the fields are
\begin{equation}
\bi{E}_{\rm L}=q\gamma \frac{Q(y)}{y^3}\bi{r}
\end{equation}

and
\begin{equation}
\bi{B}_{\rm L}=\frac{1}{c}\bi{v}\times\bi{E}_{\rm L} ,
\end{equation}
where $\bi{r}$ is the position from the centre of the particle in
the laboratory frame and $\bi{y}$ has the same meaning in the
instantaneous rest frame,
\begin{equation}
\bi{y}=\bi{r} + (\gamma-1)(\bi{r}
\cdot\hat{\bi{v}})\hat{\bi{v}} . 
\end{equation}

The local field energy and momentum are calculated integrating, over the 
volume, the energy density $u_{\rm L}=(E_{\rm L}^2+B_{\rm L}^2)/(8\pi)$
 and the momentum density
 $c^{-2}\bi{S}_{\rm L}$, $\bi{S}_{\rm L}=
c\bi{E}_{\rm L}\times\bi{B}_{\rm L}/(4\pi)$.
The results are
\begin{equation}\label{EnergyL}
U_{\rm L}=U_\rme\gamma \Big(1+\frac{1}{3}\big(\frac{v}{c}\big)^2\Big) +\Or(a)
\end{equation}
and
\begin{equation}\label{MomentumL}
\bi{P}_{\rm L}=\frac{4}{3}c^{-2}U_\rme\gamma\bi{v}+\Or(a) .
\end{equation}

If we add these results to the energy $(m_0+m_{\rm P}(1-\gamma^{-2}))c^2\gamma$
and momentum $(m_0+m_{\rm P})\gamma\bi{v}$ of the particle we obtain the
expected results for the dressed particle, $(m_0c^2+U_\rme)\gamma$ and
$(m_0+U_\rme/c^2)\gamma\bi{v}$.

 Why does the momentum of the field have the
famous $4/3$ factor? Exactly for the same reason why the momentum of the
particle is not the bare one: because the field is not free and therefore it
has stress. In the particle case, the stress tensor $P^{\mu\nu}$ is well
separated from the energy-momentum tensor $\Theta^{\mu\nu}$. The volume
integrals of the time components of the energy-momentum tensor transform
as a four-vector. The volume integrals of the time components of the stress
tensor, that is the energy and momentum contributions of stress, do not
transform as a four-vector. In the field case it is not possible in general
to split the tensor $T^{\mu\nu}$ into an energy-momentum tensor plus a
stress tensor. So the volume integrals of $T^{\mu 0}$ do not transform as
a four-vector. That is, the total energy and momentum of the field do not
form a four-vector. There is nothing wrong with this. Relativity only
requires that the \textit{total} energy and momentum of a \textit{confined}
system should form a four-vector.
 Indeed the effect of the stress of the field is cancelled by the effect
of the stress of the matter. 
In the particular case of a single rigid charged body, moving with constant
velocity, it is possible to obtain separated stress and energy-momentum
tensors of the field.
In the rest frame the $T^{00}$ is the energy-momentum tensor while the stress
is $T^{ij}$. All the above discussion is consistent with the results of
Singal \cite{Singal}.

The stress tensor $P^{\mu\nu}$ given by (\ref{StressTensor}) and
(\ref{Pressure}) is identical to the term that Schwinger \cite{Schwinger}
found that had to be added to $T^{\mu\nu}$ in order to obtain a
four-momentum with electromagnetic mass $U_\rme/c^2$.  We see that no
\textit{ad hoc} assumptions are needed, the term is just the pressure inside
the particle.

In order to completely solve the $4/3$ problem we have to prove that
(\ref{PowerBalance}) is verified.  The electrostatic force density is
\begin{equation}
\bi{f}_{\rm es}=q^2\gamma^2g(y)\frac{Q(y)}{y^3}\bi{r} .
\end{equation}
It is obvious that  $\bi{f}_{\rm es}$ does not contribute to
$\bi{F}_{\rm em}$ because $\smallint \rmd^3x\,\bi{f}_{\rm es}=0$,
but it is the leading contribution to the power unbalance. In fact
we are assuming that $\bi{f}_{\rm ex}$ does not stress the particle, so
 $\smallint \rmd^3x\,\bi{f}_{\rm ex}\cdot\delta\bi{v}=0$. On
the other hand the electric component of the
field reaction force density $\bi{f}_{\rm fr}$ is proportional to
the acceleration, therefore it gives a higher order contribution to the
power integral. Therefore
\begin{equation}
\int \rmd^3x\,\bi{f}\cdot\delta\bi{v}=
\int \rmd^3x\,\bi{f}_{\rm es}\cdot\delta\bi{v}+
\Or(a^2) .
\end{equation}
To calculate this integral we note that 
\begin{equation}
\delta\bi{v}=\frac{ \rmd\bi{r}}{\rmd t}
\end{equation}
and that $r^2=y^2+(\gamma^{-2}-1)(\bi{y}\cdot\hat{\bi{v}})^2$.
\begin{eqnarray}\label{PowerInt}
\int \rmd^3x\,\bi{f}_{\rm es}\cdot\delta\bi{v}&=&
\frac{q^2\gamma^2}{2}\int \rmd^3x\,g(y)\frac{Q(y)}{y^3}\frac{\rmd r^2}{\rmd t}
\\
&=&\frac{q^2\gamma}{2}\int \rmd^3y\,g(y)\frac{Q(y)}{y^3}\frac{\rmd}{\rmd t}
[(\gamma^{-2}-1)(\bi{y}\cdot\hat{\bi{v}})^2]
\\
&=&\frac{q^2\gamma}{6}\int \rmd y\,\frac{\rmd Q}{\rmd y}\frac{Q(y)}{y}
\frac{\rmd\gamma^{-2}}{\rmd t}
\\
&=&\frac{1}{3}U_e\frac{\rmd\gamma^{-1}}{\rmd t}
\end{eqnarray}
as expected from (\ref{PowerBalance}).

%%%%% Section V

\section{Conservation laws of the fields}
In this section we will consider the conservation laws of energy and momentum 
of the fields. If $T^{\mu\nu}$ is the momentum-energy-stress tensor of
the electromagnetic field and $f^{\mu}_{\rm em}$ is the force density
acting on the particle then the conservation law is
\begin{equation}\label{FieldLaw}
\nabla_{\mu}T^{\mu\nu}=-f^{\nu}_{\rm em} .
\end{equation}

As the electromagnetic field is the sum of local and radiation terms, 
the tensor, which is quadratic, contains 
cross terms  that are products of local and radiation fields,
$T^{\mu\nu} = T^{\mu\nu}_{\rm L} +T^{\mu\nu}_{\rm C}+T^{\mu\nu}_{\rm R}$.
  Introducing energy densities, Poynting
vectors and Maxwell tensors, (\ref{FieldLaw}) is equivalent to
\begin{equation}
\frac{\partial}{\partial t}(u_{\rm L}+u_{\rm C}+u_{\rm R})+
\nabla\cdot(\bi{S}_{\rm L}+
\bi{S}_{\rm C}+\bi{S}_{\rm R})=-\bi{f}_{\rm em}\cdot(\bi{v}+
\delta\bi{v})
\end{equation}
and
\begin{equation}
\frac{\partial}{c^2\partial t}(\bi{S}_{\rm L}+\bi{S}_{\rm C}+\bi{S}_{\rm R})
+ \nabla\cdot({\bf T}_{\rm L}+{\bf T}_{\rm C}+{\bf T}_{\rm R})=
-\bi{f}_{\rm em} .
\end{equation}

Now we will integrate both equations over the whole space. One has to be careful
because $T^{\mu\nu}_{\rm R}$ decays as $r^{-2}$. So we take a volume $V$,
surrounding the particle, so large that $T^{\mu\nu}_{\rm L}\approx 0$ and
 $T^{\mu\nu}_{\rm C}\approx 0$  outside $V$. In these conditions the
 radiated power
and the rate of radiated momentum defined as follows are independent of $V$.
\begin{equation}
P_{\rm rad}=\frac{\rmd}{\rmd t}\int_V \rmd^3x\,u_{\rm R}+
\int_{\partial V} \rmd\bi{A}\cdot\bi{S}_{\rm R} ,
\end{equation}
\begin{equation}
\bi{G}_{\rm rad}=\frac{\rmd}{c^2 \rmd t}\int_V \rmd^3x\,\bi{S}_{\rm R}+
\int_{\partial V}{\rmd}\bi{A}\cdot{\bf T}_{\rm R} .
\end{equation}

We get
\begin{equation}\label{EMpower}
\frac{\rmd}{\rmd t}(U_{\rm L}+U_{\rm C})+P_{\rm rad}=
-\bi{F}_{\rm em}\cdot
\bi{v}-\frac{U_\rme}{3}\frac{\rmd\gamma^{-1}}{\rmd t}
\end{equation}
and
\begin{equation}
\frac{\rmd}{\rmd t}(\bi{P}_{\rm L}+\bi{P}_{\rm C})+\bi{G}_{\rm rad}=
-\bi{F}_{\rm em} .
\end{equation}
Here we have used (\ref{PowerInt}) and the total cross energy and momentum
defined as
\begin{equation}
U_{\rm C}=\int \rmd^3x\,u_C
\end{equation}
and
\begin{equation}
\bi{P}_{\rm C}=\frac{1}{c^2}\int \rmd^3x\,\bi{S}_{\rm C} .
\end{equation}

The dressed particle includes, in addition to the matter momentum and the
pressure momentum, the momentum of the local fields. 
The reaction of these is $\bi{F}_{\rm L}=-\dot\bi{P}_{\rm L}$. This
reaction should
be subtracted from the self-force $\bi{F}_{\rm em}$ in order to
obtain the force that the dressed particle feels. We will call
this last force radiation reaction,
\begin{equation}
\bi{F}_{\rm R}=\bi{F}_{\rm em}-\bi{F}_{\rm L}=
\bi{F}_{\rm em}+\frac{\rmd\bi{P}_{\rm L}}{\rmd t} .
\end{equation}

Actually, $\bi{F}_{\rm R}$ contains, not only the reaction of the radiated
fields but also the reaction of the cross terms,
\begin{equation}\label{RRForceTerms}
\bi{F}_{\rm R}=-\bi{G}_{\rm rad}-\frac{\rmd\bi{P}_{\rm C}}{\rmd t} .
\end{equation}
The power of $\bi{F}_{\rm R}$ is obtained from (\ref{EMpower}) using
(\ref{EnergyL}), (\ref{MomentumL}) and the identity
 (\ref{identity}). It also contains a contribution
from the cross terms,
\begin{equation}\label{RRPowerTerms}
\bi{F}_{\rm R}\cdot\bi{v}=-P_{\rm rad}
-\frac{\rmd U_{\rm C}}{\rmd t} .
\end{equation}

Finally, the motion equation and the power equation of the dressed
particle are readily  obtained from (\ref{ParMomentum}) and
(\ref{ParEnergy}), with dressed mass $m=m_0 +U_\rme/c^2$
\begin{equation}\label{DressedForce}
m\frac{\rmd\gamma\bi{v}}{\rmd t}=\bi{F}_{\rm ex}+\bi{F}_{\rm R} ,
\end{equation}
\begin{equation}\label{DressedPower}
mc^2\frac{\rmd\gamma}{\rmd t}=(\bi{F}_{\rm ex}+\bi{F}_{\rm R})\cdot\bi{v} .
\end{equation}

%%%%% Section VI

\section{Radiation reaction}
The calculation of the self-force is much simpler if it is done in the
reference frame of instantaneous rest. To begin with, there are no
 magnetic
field contributions so we have only to calculate the electric field. It is
very convenient in this case to split the electric field into electrostatic
and  induced components, $\bi{E}=\bi{E}_\rme+\bi{E}_\rmi$, with the following
properties: $\nabla\cdot\bi{E}_\rme=4\pi\rho$, $\nabla\times\bi{E}_\rme=0$ and
 $\nabla\cdot\bi{E}_\rmi=0$. Note that $\bi{E}_\rme$ is the longitudinal and
$\bi{E}_\rmi$ the transverse field.
 With these definitions, in the rest frame,
$\bi{f}_{\rm es}=\rho\bi{E}_\rme$ and
 $\bi{f}_{\rm fr}=\rho\bi{E}_\rmi$. As we have previously explained in
the integrated self-force $\bi{F}_{\rm em}$ the electrostatic contribution
cancels out. The easiest way to determine these fields is to use the Coulomb
gauge, $\nabla\cdot\bi{A}=0$, because in this case the potential $\phi$
yields $\bi{E}_\rme$ and $\bi{A}$ yields $\bi{E}_\rmi$,
\begin{equation}
\bi{E}_\rme = -\nabla\phi
\end{equation}
and
\begin{equation}
\bi{E}_\rmi = -\frac{1}{c}\frac{\partial\bi{A}}{\partial t} .
\end{equation}

The electrostatic field is
\begin{equation}
\bi{E}_\rme(\bi{x},t) = \int \rmd^3y\,\frac{\rho(\bi{y},t)}
{|\bi{x}-\bi{y}|^3}(\bi{x}-\bi{y})
\end{equation}
and $\bi{A}$ can be obtained from the equation
\begin{equation}
\frac{1}{c^2}\frac{\partial^2\bi{A}}{\partial t^2}-\nabla^2\bi{A}=
\frac{4\pi}{c}\bi{j}+\frac{1}{c}\frac{\partial\bi{E}_\rme}{\partial t} .
\end{equation}
The solution is
\begin{equation}
\bi{A}(\bi{x},t) = \frac{1}{c}\int \rmd^3y\,\frac{\bi{j}(\bi{y},t^\prime)}
{|\bi{x}-\bi{y}|}+\frac{1}{4\pi c}\int \rmd^3y\,\frac{1}{|\bi{x}-\bi{y}|}
\frac{\partial\bi{E}_\rme(\bi{y},t^\prime)}{\partial t} ,
\end{equation}
where $t^\prime$ is the retarded time, $t^\prime=t-|\bi{x}-\bi{y}|/c$.
The induced field is therefore
\begin{equation}\label{EInduced}
\fl
\bi{E}_\rmi(\bi{x},t) = -\frac{1}{c^2}\int \rmd^3y\,\frac{1}{|\bi{x}-\bi{y}|}
\frac{\partial\bi{j}(\bi{y},t^\prime)}{\partial t}\nonumber
-\frac{1}{4\pi c^2}\int \rmd^3y\,\frac{1}{|\bi{x}-\bi{y}|}
\frac{\partial^2\bi{E}_\rme(\bi{y},t^\prime)}{\partial t^2} .
\end{equation}

For the  quasi-rigid motion the charge density is
$\rho(\bi{x},t)=q\gamma g(y)$ and the current density is
$\bi{j}=\rho(\bi{v}+\delta\bi{v})$,
where $\bi{y}=\bi{r}+(\gamma-1)(\bi{r}\cdot\hat{\bi{v}})\hat{\bi{v}}$,
$y=|\bi{y}|$ and $\bi{r}=\bi{x}-\bi{x}_0(t)$.
With these definitions we have in the rest frame
\begin{equation}
\frac{\partial^2\rho(\bi{x},t)}{\partial t^2}=-q\frac{\rmd g}{\rmd r}
\hat{\bi{r}}\cdot\bi{a}(t)+\Or(a^2)
\end{equation}
and
\begin{equation}\label{jDer}
\frac{\partial\bi{j}(\bi{x},t)}{\partial t}= q g(r)\bi{a}(t)
+\Or(a^2) .
\end{equation}

If one now replaces the expression for $\rho$ into the electrostatic field
equation one gets
\begin{equation}\label{Der2Ee}
\frac{\partial^2\bi{E}_\rme(\bi{x},t)}{\partial t^2} =
 q\frac{\partial}{\partial\bi{x}} \int d^3y\,\frac{\rmd g}{\rmd y}
\frac{\hat{\bi{y}}\cdot\bi{a}(t)}{|\bi{x}-\bi{y}|} .
\end{equation} 
The integral in (\ref{Der2Ee}) can be evaluated,
\begin{equation}\label{Integral}
\int \rmd^3y\,\frac{\rmd g}{\rmd y}\frac{\hat{\bi{y}}}{|\bi{x}-\bi{y}|}=
-\frac{Q(x)}{x^2}\hat{\bi{x}} ;
\end{equation}
here $Q(x)$ is defined in (\ref{Q}) and we have used the fact that
\begin{equation}
\frac{1}{4\pi}\int \rmd\Omega_y\,\frac{\hat{\bi{y}}}{|\bi{x}-\bi{y}|}=
\frac{{\rm min}(x,y)}{3{\rm max}(x,y)^2}\hat{\bi{x}} .
\end{equation}

The second derivative of the electrostatic field is obtained by taking the
gradient of (\ref{Integral}), then
\begin{equation}
\frac{\partial^2\bi{E}_\rme(\bi{x},t)}{\partial t^2} =
 -q\Big[4\pi g(x)\hat{\bi{x}}\hat{\bi{x}}+ \frac{Q(x)}{x^3}
({\bf I} -3\hat{\bi{x}}\hat{\bi{x}})\Big]\cdot\bi{a}(t) .
\end{equation}
Here ${\bf I}$ is the identity tensor. This expression should be replaced
 into (\ref{EInduced}) and
then integrated again in order to obtain the self-force
\begin{equation}
\bi{F}_{\rm em}(t)=q\int \rmd^3x\,g(x)\bi{E}_\rmi(\bi{x},t) .
\end{equation}
In this process the induced field is spherically averaged. Because of
(\ref{jDer})  and as
\begin{equation}
\frac{1}{4\pi}\int \rmd\Omega_x\,\hat{\bi{x}}\hat{\bi{x}} =
\frac{1}{3}{\bf I} ,
\end{equation}
one finds that the contribution of the second term of (\ref{EInduced})
is exactly $-1/3$ of that of the first term.
Finally we obtain the self-force formula in the rest frame
\begin{equation}\label{SelfForce}
\bi{F}_{\rm em}(t)=
-\frac{2}{3}\frac{q^2}{c^2}\int\!\!\!\int \rmd^3x\,\rmd^3y\,\frac{g(x)g(y)}
{|\bi{x}-\bi{y}|}\bi{a}(t-|\bi{x}-\bi{y}|/c) .
\end{equation}
As was expected it is an average of the retarded accelerations, therefore no
violation of causality is due to this formula. By using the Coulomb gauge
 we have
obtained $\bi{F}_{\rm em}$ in closed form instead of the troublesome
power expansion one gets when the Lorenz gauge is used. The sum of that power
expansion appears in the textbook of Jackson \cite{Jackson}, and indeed our
equation (\ref{SelfForce}) can be deduced from equations (16.28--16.30) of
that book. 

When the acceleration is constant the self-force reduces to the
reaction of the local fields,
\begin{equation}
\bi{F}_{\rm L}(t)=-\frac{4}{3}\frac{U_e}{c^2}\bi{a}(t) .
\end{equation}
One has to subtract this contribution in order to obtain the radiation
reaction force
\begin{equation}\label{RRForce}
\bi{F}_{\rm R}(t)=
\frac{2}{3}\frac{q^2}{c^2}\int\!\!\!\int \rmd^3x\,\rmd^3y\,\frac{g(x)g(y)}
{|\bi{x}-\bi{y}|}\big(\bi{a}(t)-\bi{a}(t-|\bi{x}-\bi{y}|/c)\big) .
\end{equation}

If one expands the retarded acceleration in terms of $|\bi{x}-\bi{y}|$
the LAD result is recovered as the first term in the expansion,
\begin{equation}
\bi{F}_{\rm R}=\frac{2}{3}\frac{q^2}{c^3}\dot\bi{a}(t)+\Or(R) .
\end{equation}

If one uses (\ref{SelfForce}) to calculate the self-force of
a  charged spherical shell of radius $R$ the
Sommerfeld-Page-Caldirola-Yaghjian result is obtained,
\begin{equation}\label{Caldirola}
\bi{F}_{\rm em}(t)=-\frac{1}{3}\frac{q^2}{cR^2}\big(\bi{v}(t)
-\bi{v}(t-2R/c)\big) .
\end{equation}

%%%% Section VII

\section{Integral equation and point particle equation}
In the rest frame the equation of motion is
\begin{equation}\label{NonRelEq}
m\bi{a}=\bi{F}_{\rm ex}+\bi{F}_{\rm R} .
\end{equation}
This is, of course, the non-relativistic equation.  The condition
 $a\ll c^2/(2R)$
implicit in the quasi-rigid approximation implies that $v\ll c$ for
times of the order of the transit time (\textit{i.~e.} the time that the light
takes to traverse the particle) and therefore the non-relativistic
equation can
be used. Our purpose is to solve (\ref{NonRelEq}) in order to express
the radiation reaction as a functional of the external force, instead of
a functional of the acceleration. Equation (\ref{RRForce}) can be written
as
\begin{equation}\label{RRKernel}
\bi{F}_{\rm R}(t)=\int_0^\infty \rmd t^{\prime}\,k(t^{\prime})\big
(\bi{a}(t)-\bi{a}(t-t^{\prime})\big)
\end{equation}
where the kernel k is given by
\begin{equation}\label{kernel}
k(t)=\frac{2q^2}{3c^2}\int\!\!\!\int \rmd^3x\,\rmd^3y\,
\frac{g(x)g(y)}{|\bi{x}-\bi{y}|}\delta(t-|\bi{x}-\bi{y}|/c) .
\end{equation}

The solution is obtained by means of a Fourier transform
\begin{equation}\label{K}
K(\omega)=\int_0^\infty \rmd t\, k(t)\rme^{\rmi\omega t}
=\frac{2q^2}{3c^2}\int\!\!\!\int \rmd^3x\,\rmd^3y\,
\frac{g(x)g(y)}{|\bi{x}-\bi{y}|}\rme^{\rmi\omega|\bi{x}-\bi{y}|/c} .
\end{equation}

 The  mass of the local fields equals $K(0)$,
\begin{equation}\label{K0}
K(0)=\int_0^\infty \rmd t\, k(t)
=\frac{2}{3}\frac{q^2}{c^2}
\int\!\!\!\int \rmd^3x\,\rmd^3y\,\frac{g(x)g(y)}{|\bi{x}-\bi{y}|}=
\frac{4}{3c^2}U_\rme\ ;
\end{equation}
while the derivative of $K(\omega)$ at $\omega=0$ gives the constant of
the LAD equation,
\begin{equation}\label{DerK0}
-\rmi\frac{\rmd K}{\rmd\omega}\bigg\vert_{0}=\int_0^{\infty}\rmd t\,tk(t)
= \frac{2}{3}\frac{q^2}{c^3}\int\!\!\!\int \rmd^3x\,\rmd^3y\,g(x)g(y)=
\frac{2}{3}\frac{q^2}{c^3} .
\end{equation}

From (\ref{K0}) and (\ref{DerK0}) one obtains an expression
for the average delay time $\tau$
\begin{equation}\label{tau}
\tau=\frac{1}{K(0)}\int_0^{\infty}\rmd t\,tk(t)
=\Bigg(c\int\!\!\!\int \rmd^3x\,\rmd^3y\,\frac{g(x)g(y)}{|\bi{x}-\bi{y}|}
\Bigg)^{-1} .
\end{equation}

The acceleration, obtained from (\ref{NonRelEq}), is then
\begin{equation}\label{Acceleration}
\bi{a}(t)=\int_{-\infty}^{\infty}\rmd t^{\prime}\,G(t^{\prime})
\bi{F}_{\rm ex}(t-t^{\prime}) ,
\end{equation}
where
\begin{equation}
G(t)=\frac{1}{2\pi}\int_{-\infty}^{\infty}d\omega\,
\frac{\rme^{-\rmi\omega t}}{m-K(0)+K(\omega)} .
\end{equation}

Causality requires that $G(t)=0$ for $t<0$. That implies that
$(m-K(0)+K(\omega))^{-1}$ must be regular for ${\rm Im}(\omega)>0$.
$K(\omega)$ is regular for ${\rm Im}(\omega)>0$, so the only possible
singularities are poles when $m-K(0)+K(\omega)=0$. 
For $\eta \geq 0$ $K(\rmi\eta)$ is real and $\lim_{\eta\to\infty}K(\rmi\eta)=0$.
When $g(r)\geq 0$, $K(\rmi\eta)$ decreases monotonically from  $K(0)$ to zero.
Even when there are regions in which $g(r)$ is negative $K(\rmi\eta)> 0$.
 Therefore
 $(m-K(0)+K(\omega))^{-1}$ is regular in the upper half of the complex
plane if, and only if,
$m>K(0)$. That is, it must be $m>\frac{4}{3}U_\rme/c^2$, or equivalently
$m_0>\frac{1}{3}U_\rme/c^2$. Note that this is the same condition discussed
in section 3.  If the condition is fulfilled
$G(t)\sim \theta(t)\exp(-t/\tau_G)$ and $\tau_G\sim \tau$.
 If the condition is not satisfied, causality
is violated and run-away solutions appear. Let us look in more detail how
this happens. Consider a case in which the external force was constant
for $t<0$, and then suddenly vanishes at $t=0$. The self-force
 $\bi{F}_{\rm em}$ given by (\ref{SelfForce})  is opposite
to the acceleration and remains the same at the transition,
$\bi{F}_{\rm em}(0^{+})=\bi{F}_{\rm em}(0^{-})=-K(0)\bi{a}(0^{-})$.
 The motion is determined by (\ref{ParMomentum}). For $t>0$ the
acceleration  equals $\bi{F}_{\rm em}$ divided by the
effective mass of matter $m_0+m_{\rm P}=m-K(0)$, in particular
$\bi{a}(0^{+})=-(m-K(0))^{-1}K(0)\bi{a}(0^{-})$.
If $m_0+m_{\rm P}>0$ the new acceleration is opposite to that of negative
times and,
as time elapses, the accelerations for $t>0$ subtract from the previous ones
in (\ref{SelfForce}); therefore the self-force is reduced and the
acceleration  vanishes exponentially.
On the contrary if $0<m<K(0)$, the new acceleration not only has the same
sense of the previous one but it is actually larger than it. As time elapses the
self-force increases and the acceleration diverges exponentially.

 The mass condition
was previously demonstrated for  the spherical charged shell by Moniz and 
Sharp \cite{Moniz}. Although their quantum-mechanical results have been
disputed \cite{Grotch}, their classical calculation of the run-away solutions
is correct, apart from the fact that they do not include the pressure
contribution to the mechanical mass. 
Here we have shown that the mass condition is valid for any charge distribution
and have clarified its relation with the inertia of stress.

The equation (\ref{Acceleration}) can be re-written as
\begin{equation}\label{Acceleration2}
\bi{a}(t)=\bigg[\int_0^{\infty}\rmd t^{\prime}\,G(t^{\prime})\bigg]
\bi{F}_{\rm ex}(t)
+\int_0^{\infty}\rmd t^{\prime}\,G(t^{\prime})\big(
\bi{F}_{\rm ex}(t-t^{\prime})-\bi{F}_{\rm ex}(t)\big) .
\end{equation}
The pre-factor of the force in the first term of (\ref{Acceleration2})
happens to be $m^{-1}$
\begin{equation}
\int_0^{\infty}\rmd t\,G(t)=\int_{-\infty}^{\infty}\rmd\omega\,
\frac{\delta(\omega)}{m-K(0)+K(\omega)}=\frac{1}{m} .
\end{equation}

The radiation reaction is then
\begin{equation}\label{RRForceInt}
\bi{F}_{\rm R}=m\int_0^{\infty}\rmd t^{\prime}\,G(t^{\prime})\big(
\bi{F}_{\rm ex}(t-t^{\prime})-\bi{F}_{\rm ex}(t)\big) .
\end{equation}

Finally, if the external force varies slowly enough, more precisely if
\begin{equation}\label{PointChargeCondition}
\frac{|\dot F_{\rm ex}|}{F_{\rm ex}}\ll \tau^{-1}
\end{equation}
an approximated instantaneous radiation reaction is obtained by taking
the first term of the expansion in $t^{\prime}$ of the retarded force in
 (\ref{RRForceInt}),
\begin{equation}\label{RRForceDer}
\bi{F}_{\rm R}=\frac{2}{3}\frac{q^2}{c^3 m}\dot\bi{F}_{\rm ex} .
\end{equation}
To obtain this we have used the fact that
\begin{eqnarray}
\int_0^{\infty}\rmd t\,tG(t)&=&\rmi\int_{-\infty}^{\infty}\rmd\omega\,
\frac{1}{m-K(0)+K(\omega)}\frac{\rmd\delta(\omega)}{\rmd\omega}
\nonumber\\
&=& \frac{\rmi}{m^2}\frac{\rmd K}{\rmd\omega}\bigg\vert_0
=-\frac{2}{3}\frac{q^2}{c^3 m^2} .
\end{eqnarray}

The equation (\ref{RRForceDer}) was proposed by Rohrlich \cite{Rohrlich3}
as the correct one for point particles. We see that as long as the mass
condition $m>\frac{4}{3}U_\rme/c^2$ is fulfilled the radiation reaction can
be expressed by (\ref{RRForceInt}) which conforms to causality and does
not produce run-away solutions. The approximated expression of
 (\ref{RRForceDer}) does produce violations of causality when
$F_{\rm R}\sim F_{\rm ex}$. When this happens the condition of
 (\ref{PointChargeCondition}) is not satisfied, so the approximation
is not valid.

A particle can be considered a point particle if its radius is negligible
when compared with the wave-length  of the radiated field. This condition
is equivalent to (\ref{PointChargeCondition}). So 
(\ref{RRForceDer}) can indeed be considered the correct one for
point particles. It is worth noting that applied forces with discontinuities,
that violate (\ref{PointChargeCondition}), are very common. However,
usually such discontinuities are not real, but the result of some 
approximation. In most cases the formula of (\ref{RRForceDer}) behaves
properly if, instead of discontinuities,  realistic smoothed steps are used.

%%%%% Section VIII

\section{Radiation reaction four-force and radiated power}
The radiation reaction of the physical point charge with $\bi{v}=0$
is given by equation (\ref{RRForceDer}). In this section we find this
force for a particle moving with arbitrary speed.

A four-force $K^{\mu}$ that conserves mass is related to the force by
$K^i=\gamma F^i$ and to the power by $K^0=\gamma\bi{F}\cdot\bi{v}/c$.
This implies that it is orthogonal to the four-velocity $K^{\mu}u_{\mu}=0$.
 In the rest frame $K^i=F^i$ and $K^0=0$.

 The motion equations of the dressed mass,
(\ref{DressedForce}) and (\ref{DressedPower}), show that the radiation
reaction is a mass-conserving force. Then, the four-force of the radiation
reaction of the physical point charge is in the rest frame
\begin{equation}
K^i_R = \frac{2}{3}\frac{q^2}{c^3 m}\frac{\rmd F^i_{\rm ex}}{\rmd t}
\end{equation}
and
\begin{equation}
K^0_R=0 .
\end{equation}

Therefore, if we call $K^{\mu}$ the external four-force and $\tau$ the proper
time, the radiation reaction four-force is in an arbitrary reference frame
given by
\begin{equation}\label{RRFourForce}
K^{\mu}_R = \frac{2}{3}\frac{q^2}{c^3 m}\Bigr[\frac{\rmd K^{\mu}}{\rmd\tau}
+\frac{1}{c^2}\big(\frac{\rmd K^{\nu}}{\rmd\tau}u_{\nu}\big)u^{\mu}\Big] .
\end{equation}

The four-scalar in parenthesis is
\begin{equation}
\frac{\rmd K^{\mu}}{\rmd\tau}u_{\mu}=-\gamma^3\bi{F}_{\rm ex}\cdot\bi{a} .
\end{equation}

From (\ref{RRFourForce}) the force and power expressions are obtained
for any speed.
\begin{equation}\label{RRForceRel}
\bi{F}_{\rm R}= \frac{2}{3}\frac{q^2}{c^3 m}\Bigr[
\frac{\rmd\gamma\bi{F}_{\rm ex}}{\rmd t}-
\frac{\gamma^3}{c^2}(\bi{F}_{\rm ex}\cdot\bi{a})\bi{v}\Big]
\end{equation}
and
\begin{equation}\label{RRPowerRel}
\bi{F}_{\rm R}\cdot\bi{v}= \frac{2}{3}\frac{q^2}{c^3 m}\Bigr[
\frac{\rmd\gamma\bi{F}_{\rm ex}\cdot\bi{v}}{\rmd t}-
\gamma^3\bi{F}_{\rm ex}\cdot\bi{a}\Big] .
\end{equation}
From these equations, and equations (\ref{RRForceTerms}) and
 (\ref{RRPowerTerms})
one obtains the momentum and energy of the cross terms, the radiated power,
and rate of radiated momentum,
\begin{equation}
\bi{P}_{\rm C}= -\frac{2}{3}\frac{q^2}{c^3m}\gamma\bi{F}_{\rm ex} ,
\end{equation}
\begin{equation}
U_{\rm C}= -\frac{2}{3}\frac{q^2}{c^3 m}
\gamma\bi{F}_{\rm ex}\cdot\bi{v}=
\bi{P}_{\rm C}\cdot\bi{v} ,
\end{equation}

\begin{equation}\label{RadPower}
P_{\rm rad}= \frac{2}{3}\frac{q^2}{c^3 m}
\gamma^3\bi{F}_{\rm ex}\cdot\bi{a} ,
\end{equation}
\begin{equation}
\bi{G}_{\rm rad}= \frac{2}{3}\frac{q^2}{c^3m}
\frac{\gamma^3}{c^2}(\bi{F}_{\rm ex}\cdot\bi{a})\bi{v}=
\frac{1}{c^2}P_{\rm rad}\bi{v} .
\end{equation}

Equation (\ref{RadPower}) reduces to the Larmor result \cite{Larmor}
 in the non-relativistic limit only if
 $\dot{\bi{F}}_{\rm ex}\cdot\bi{a}=0$. The Larmor result is valid
for a mathematical point charge ($R=0$), not for an extended particle.

%%%%%% Section IX

\section{Conclusion}
We have completely solved the self-force of the extended quasi-rigid
particle. We have shown that, provided $m_0>\frac{1}{3}U_\rme/c^2$, it is a
perfectly consistent classical system
 conforming to causality and conservation
of energy and momentum. In order to obtain the correct equations it is
essential to use the {\it relativistic} conservation equations of energy and
momentum for the continuous medium,  even if $v\ll c$; only so the inertia
of pressure is properly taken into account. The mathematical point charge
($R=0$) is inconsistent with classical electrodynamics. For a given dressed
mass $m$ and charge $q$, the minimum radius that any particle can have is
$(2q^2)/(3mc^2)$. Nevertheless if the radius of a particle is much smaller
than any other distance in the problem, the radius becomes irrelevant and the
particle can be properly considered as a point charge from the physical
point of view.

 A comment about QED. A correct quantum theory of a system,
should give in the classical limit a proper classical system. As the
mathematical point charge is inconsistent in classical electrodynamics,
it follows that the proper quantum theory
of a structureless particle (such as the electron) should have as  classical
limit an {\it extended} classical particle. At first sight one might think
that this were not impossible as the
single particle picture breaks down at the Compton length scale, which is
$\alpha^{-1}$ ($\approx 137$) times larger than the classical radius of the
particle. The relativistic quantum non-localities produce an effective
particle extension. Nevertheless
the QED calculation for spin $\frac{1}{2}$ of Low \cite{Low} shows that
these effects are not enough to eliminate the run-away solutions, but that
the size bound is reduced by a factor of $\exp(-\alpha^{-1})$.

%%%% Acknowledgment
\ack
I wish thank Dr. Victor Villalba for many enlightening discussions and
for thoroughly reading this piece of work.

\vfill
%%%%%% References

% Referencias para  Journal of Physics A

\end{document}